\documentclass[a4paper,11pt]{article}
\pdfoutput=1 

\usepackage{jinstpub} 

\title{\boldmath Performance of the SDHCAL technological prototype}

\author[a]{G. Grenier}
\affiliation{IPNL/CNRS-IN2P3/Universit\'e Lyon~1, \\ 
4, rue Enrico Fermi, 69622 Villeurbanne Cedex, France}

\emailAdd{grenier@ipnl.in2p3.fr}

\abstract{The SDHCAL technological prototype is a 
$1 \times 1 \times 1.3$~m$^3$ high-granularity Semi-Digital Hadronic 
CALorimeter using Glass Resistive Plate Chambers as sensitive medium. 
It  is one of the two HCAL 
options considered by the ILD Collaboration to be proposed
for the detector of the future International Linear Collider project.
The prototype is made of up to 50 GRPC detectors of 1~m$^2$ size and 
3~mm thickness each with an embedded semi-digital electronics readout
that is autotriggering and power-pulsed. The GRPC readout is finely segmented into pads of 1~cm$^2$.
This proceeding describes the prototype, its operation and its performance in energy reconstruction.
Aspects of the GRPC readout modelling and comparisons with simulations are also presented.}


\arxivnumber{1605.00418} 

\collaboration[c]{On behalf of CALICE collaboration}

\proceeding{13$^{\text{th}}$ Workshop on Resistive Plate Chamber and Related Detectors\\
  February 22-26, 2016\\
  Universiteit Gent, Gent, Belgium}

\begin{document}
\graphicspath{{images/}}
\maketitle
\flushbottom

\section{Introduction}\label{sec:intro}
This article describes some performances of the Semi-Digital Hadronic CALorimeter (SDHCAL) technological prototype~\cite{Baulieu:2015pfa}. 
The SDHCAL is one of the two hadronic calorimeter options considered by the 
ILD (International Large Detector) Collaboration~\cite{Behnke:2013lya} 
to be proposed for the detector of the future 
International Linear Collider (ILC). ILC detectors are designed for 
Particle Flow Algorithms~(PFA)~\cite{Brient:2004yq}. For optimal use 
of PFA, calorimeters need to be homogeneous and finely segmented. 

\section{SDHCAL concept}\label{sec:sdhcal}

The SDHCAL prototype meets the ILC requirements by the combination of 
various technological choices. The homogeneity is achieved by the use 
of large Glass Resistive Plate Chambers (GRPC) as the active 
medium combined with a power-pulsed embedded electronics. 
The power-pulsing suppresses the need of integrating a cooling circuit 
inside the detector by reducing the power consumption and heating. 
The homogeneity is further achieved by having all services 
(gas inflows and outflows, high voltage, data readout, etc.) 
only on one side of the GRPC and outside the HCAL. 

The SDHCAL prototype is a sampling calorimeter with 
2~cm thick absorber layers, nearly one radiation length, 
and 6~mm thick active detectors.
The prototype consists of a self-supporting mechanical structure~\cite{Baulieu:2015pfa} made of absorber plates. The structure is designed to be able to hold up to 50 GRPC cassettes. These cassettes contain the sensitive elements and their wrapping contributes to the absorber thickness.

\section{The GRPC cassette}\label{sec:cassette}

\begin{figure}[tbp] 
\begin{center}
\includegraphics[width=.95\textwidth]{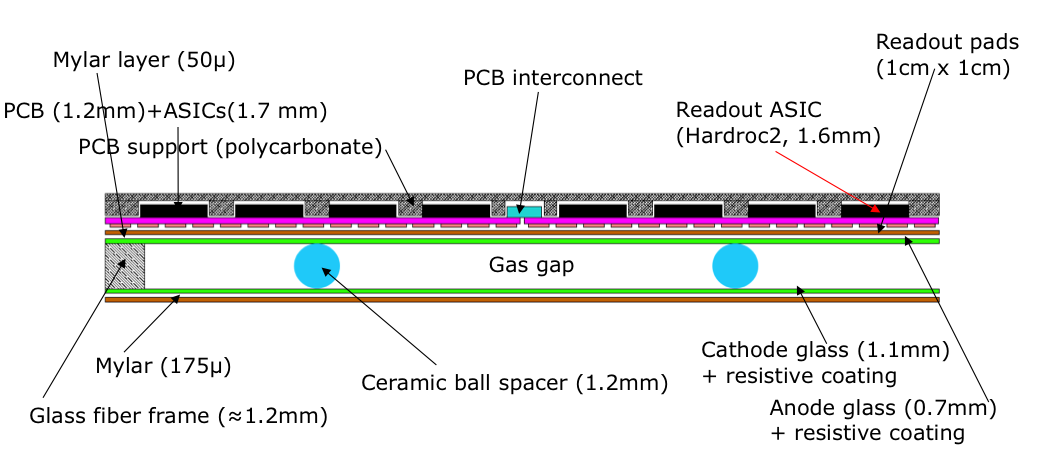}
\caption{Schematic cross-section view of a 1~m$^2$ GRPC cassette.} 
\label{fig:SDHCAL:Cassette} 
\end{center}
\end{figure}
 
A GRPC cassette~\cite{Baulieu:2015pfa} has a 1~m$\times$1~m area 
and is 11~mm thick (see figure~\ref{fig:SDHCAL:Cassette}).
It contains one GRPC and its associated electronics. 
The cassette is a thin box consisting of two 2.5~mm thick 
stainless steel plates separated by 6~mm wide stainless steel 
spacers which form the walls of the box.
One of the 2 plates has a 1~m$\times$1.2~m area. The extra 20~cm are used to hold 
the PCBs used for the data acquisition as well as the gas outlets 
and the high voltage box. 
Precision machined stainless steel spacers, insulated from the GRPC, 
are making the cassette's sides. A polycarbonate mask is added around 
the ASICs to ensure that once the cassette is closed, the PCB is 
forced to stay into contact with the GRPC anode. This cassette 
structure ensures a homogeneous efficiency of the GRPC signal 
collection by the PCB copper pads.

\subsection{GRPC}\label{sec:GRPC}
The GRPC is used in saturated avalanche mode~\cite{Ammosov:2007zz}:  
the avalanche is initiated by the crossing of the 1.2~mm thick 
gas gap by one or more charged particles. The gap is framed by two 
electrodes made of borosilicate float glass~\cite{Bedjidian:2010zz}. 
The anode and cathode thicknesses are 0.7~mm and 1.1~mm respectively. 
The smaller anode thickness enhances the signal in the copper 
pad closest to the crossing particle and lowers the relative signal 
seen by neighbouring pads. The high voltage used is typically 
7~kV. A glass fibre frame, width 3~mm, height 1.2~mm, 
is used to seal the gas volume. 
The operating gas is a mixture of 93\% of TFE, 5\% of C0$_2$ and 
2\% of SF$_6$. TFE has been chosen for its low ionisation energy 
enabling efficient creation of avalanches. 
CO$_2$ and SF$_6$ are used as UV and electron quenchers respectively.
Comparisons with other gas mixtures are described in \cite{Kieffer:2011zea}.

\subsection{Electronics}\label{sec:electronics}
A 1~m$^2$~GRPC is tiled with 6 PCB of size of 
$\frac{1}{3} \times \frac{1}{2}$~m$^2$. The PCBs have eight layers.
On one external face, 1536 copper pads of $1\times 1$~cm$^2$ are 
printed. Copper pads are separated by 406~$\mu$m.
On the opposite face, 24 HARDROC2 ASICs~\cite{Callier:2014uqa} are 
soldered. Each ASIC is connected to $8\times 8$ pads through 
the PCB. The electronic channel cross-talk between two adjacent pads 
is less than 2\%~\cite{Bedjidian:2010yj}. 
The readout electronics system then achieves the SDHCAL fine transversal segmentation.

Each of the 64 channels of an ASIC is read by a semi-digital readout (2-bit, 3 discriminators)
covering a total channel dynamic range from 10~fC to 15~pC. 
The typical size of an avalanche inside the GRPC is 
around 1~mm$^2$. At high energy, the shower core is very dense and the gas gap can be crossed by more than 100 particles per cm$^2$. 
A simple binary readout will suffer from saturation effect.
Semi-digital readout reduces this effect and 
improves the energy reconstruction.

When the ASIC is in acquisition mode, the 192 discriminator outputs 
are checked each 200~ns. If one is fired, an event is stored in an 
integrated digital memory. This auto-triggering mechanism enables the 
operation of the calorimeter without external triggers. An ASIC event 
consists of the ASIC identification number, a 3-byte clock counter
and the output state of the 192 discriminators. The HARDROC2 can store 
up to 127 ASIC events. Once the memory is full, the ASIC
raises a signal and waits for instruction from the external Data 
Acquisition system.

\section{Prototype operation}\label{sec:operation}

The full prototype has been exposed 
to positive (SPS H2 line) and negative (SPS H6 line) pion and electron 
beams at the CERN SPS for 3 periods totalling 5 
weeks. 
The beam optics were set to enlarge the beam and the particle rate was set to a maximum
of 1000 particles per spill. The spill length was 9 seconds. The beam energy was ranging between 5 and 80 GeV.

For these beam tests, 48 GRPCs were inserted inside the 
absorber structure. The 3 thresholds for the 2-bit readout were set to 
114 fC, 5 pC and 15 pC respectively, the average MIP induced charge 
being around 1.2 pC. 
The same electronics gain was used for all the channels and the 
high voltage applied on the GRPC was 6.9 kV. The proportion of dead 
channels was about  0.1\%. 

\section{Data analysis}\label{sec:analysis}
\subsection{Time analysis}\label{sec:time}
In the trigger-less data acquisition mode used for the test beam, 
all the activity in the detector is recorded: 
isolated hits (fired pads) due to noise, particles crossing the detector or 
showering in it. The electronics runs with a clock tick of 200~ns.
For the vast majority of clock ticks, there are no hits 
recorded in the prototype. Analysis of the random hit distribution
leads to a mean noise estimation of 0.35 hits per 200~ns for 
the more than 460000 readout channels~\cite{Baulieu:2015pfa}. 
This observed noise rate is averaged over all ASICs and 
has negligible impact for hadronic calorimetry: 
hadronic showers usually produce hundreds of hits. 
Local variation of the gap, of the painting resistivity or of the temperature may 
lead to hot spots where the noise can reach hundreds of Hz.  A few ASICs
($ \simeq $ 2 \%) located on the edge of very few  chambers reach up to 4 kHz. The reason behind such behavior was found to be  a slightly reduced frame height leading to a locally increased electric field~\cite{Baulieu:2015pfa}. 
Excluding hot spots, the noise rate is found to be of the order of 0.1 Hz/cm$^2$.

Potential physics events are reconstructed by combining hits 
from 3 consecutive clock ticks if the middle clock tick has at 
least 7 hits. 
To monitor the calorimeter performance, the efficiency and 
multiplicity of the GRPC response are estimated using beams of muons.
The efficiency has been measured to be around 96\% for each GRPC. 
The mean multiplicity, i.e. the mean number of fired pads per crossing particle, 
has been measured to be 1.7 hits~\cite{Buridon:2016ill}.

\subsection{Pion energy reconstruction}
A simple selection of showers due to interacting pions is done for energy reconstruction: 
electrons are rejected by requiring that the shower starts after the
fifth layer or that more than 30 layers have at least 4 hits. Muons
are rejected by asking that the mean number of hits per fired layers 
should be above 2.2 and that at least 20\% of the fired layers
have a spatial hit distribution with a RMS above 5~cm. Neutral 
particles are rejected by requiring the presence of at least 4 hits
in the first 5 layers. Details of the selection are described 
in~\cite{Buridon:2016ill}.

Though the beam conditions were set to a low particle rate, the beam intensity was too high for GRPC full recovery time. 
A correction on the number of hit in showers as a function of the time $t$ since the spill starts have been implemented. This correction is linear in $t$ for hadronic showers~\cite{Buridon:2016ill} and cubic-polynomial for electromagnetic showers~\cite{Deng:2016obt}.
The correction is estimated from the data separately for each of the 3 readout thresholds.  
The three thresholds are used to tag pads fired by few, 
little or many crossing particles inside the shower.

The energy of a pion is reconstructed as 
\begin{equation}
\label{eq:Ereco}
E_{reco}=\alpha(\mathrm{N}_{\mathrm{hit}}) \mathrm{N}_{\mathrm{hit} 1} + \beta(\mathrm{N}_{\mathrm{hit}}) \mathrm{N}_{\mathrm{hit} 2} + \gamma (\mathrm{N}_{\mathrm{hit}}) \mathrm{N}_{\mathrm{hit} 3}
\end{equation}
where $\mathrm{N}_{\mathrm{hit}}=\mathrm{N}_{\mathrm{hit} 1}+\mathrm{N}_{\mathrm{hit} 2}+\mathrm{N}_{\mathrm{hit} 3}$
and, $\alpha$, $\beta$ and $\gamma$ are quadratic 
functions. 
$\mathrm{N}_{\mathrm{hit} j}$ is the time-corrected number of hits for which the $j^{th}$ threshold is the highest crossed threshold.
The calorimeter response
is linear with deviations from linearity below 5\%
for pion energies between 7~GeV and 80~GeV. The corresponding resolution 
decreases from 25\% for 5~GeV pions down to 7.7\% for 80~GeV 
pions~\cite{Buridon:2016ill}.

\section{Simulation}

A GEANT4~\cite{Agostinelli:2002hh} based simulation for the prototype has been implemented. 
A dedicated algorithm simulating the GRPC electronic response to a charged particle crossing has been developed and tuned on data~\cite{Deng:2016obt}.
This simulated response is obtained for one GRPC by looping on all charged particles crossing the chamber. For each of these particles, the induced charge is simulated to reproduce efficiency for isolated muon data. This charge is then dispatched on the neighbouring pads. The dispatching parameters are tuned to reproduce the muon multiplicity. After all induced charges have been simulated, the final step is to apply the thresholds.

\subsection{Efficiency modelling}
\begin{figure}[tbp]
\centering
\mbox{\includegraphics[width=.45\textwidth,trim=0 1cm 0 9cm, clip=true]{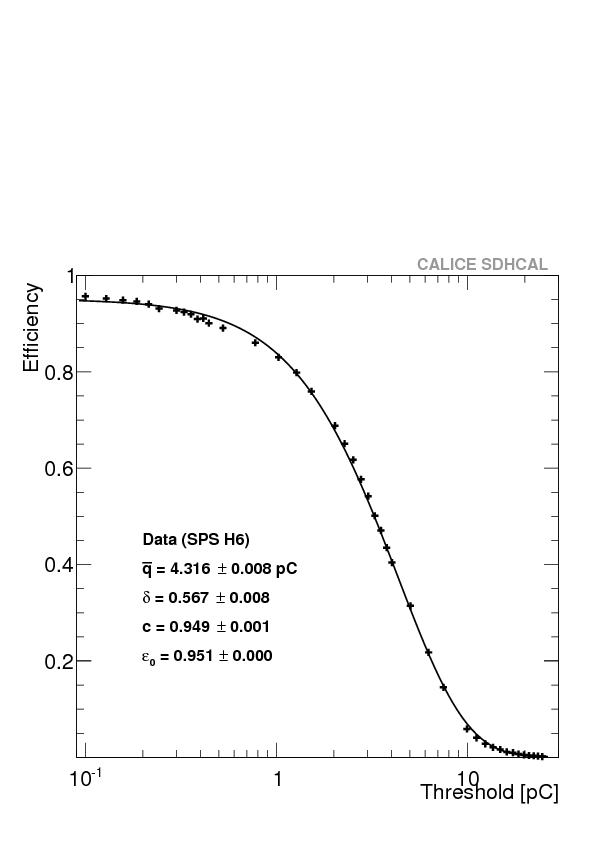}}
\mbox{\includegraphics[width=.45\textwidth,trim=9cm 1cm 0 0, clip=true]{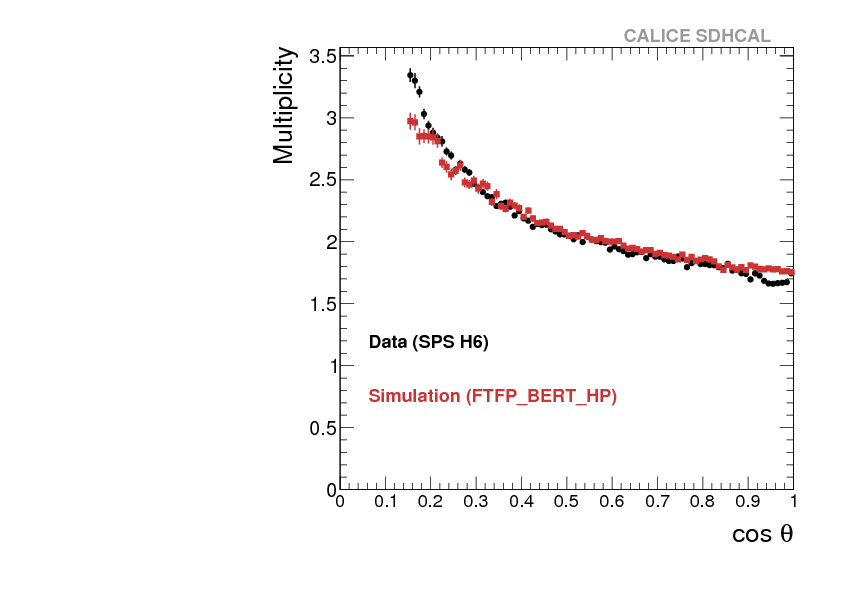}}
\caption{Left: threshold scan results with muon beams: average efficiency as a function of threshold for data (points) and modelling (solid line). Right: average pad multiplicity as a function of $\cos\theta$ with black circles and red squares for data and simulation respectively. $\theta$ is the incident angle of the muon crossing the GRPC.}
\label{fig:muonmodel}
\end{figure}

During special runs, the thresholds were varied in a few layers 
to measure the efficiency as a function of the threshold~\cite{Deng:2016obt} as 
shown in figure~\ref{fig:muonmodel} left. The  charge induced by an 
avalanche can be modelled by a Polya distribution 
(eq.~\ref{eq:polya}) and eq. 9 of \cite{Fonte:2012hf}. 
The efficiency is modelled as the combination 
of the probability $\epsilon_0$ for a crossing
charged particle to initiate an avalanche and the probability that 
the induced charge is above the threshold (eq.~\ref{eq:effvsthreshold}). 
\begin{eqnarray}
\label{eq:polya}
P(q)&=&\frac{1}{\Gamma(1+\delta)}\big(\frac{1+\delta}{\bar q}\big)^{1+\delta}q^\delta e^{[-\frac{q}{\bar q}(1+\delta)]}\\
\label{eq:effvsthreshold}
\varepsilon(q)&=&\varepsilon_0 - c\int_0^q P(q') dq' 
\end{eqnarray}

\subsection{Multiplicity modelling}
To model the muon pad multiplicity, the simulated induced charge is 
increased by a factor related to the length of the charged particle travel in the gas gap.
This correction is tuned to reproduce the angular dependence of the pad multiplicity as shown on figure~\ref{fig:muonmodel} right~\cite{Deng:2016obt}.
The corrected induced charge spread on the anode is modeled by the sum of two 2D Gaussian distributions and each pad is given the fraction of the induced charge in front of it.

\subsection{Shower modelling} 
\begin{figure}[tbp]
\centering
\mbox{\includegraphics[width=.45\textwidth]{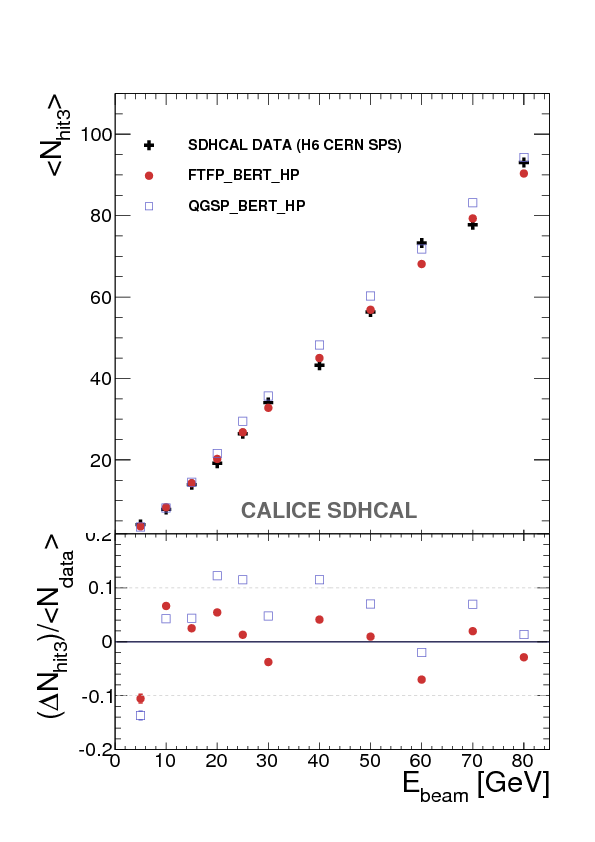}}
\mbox{\includegraphics[width=.45\textwidth]{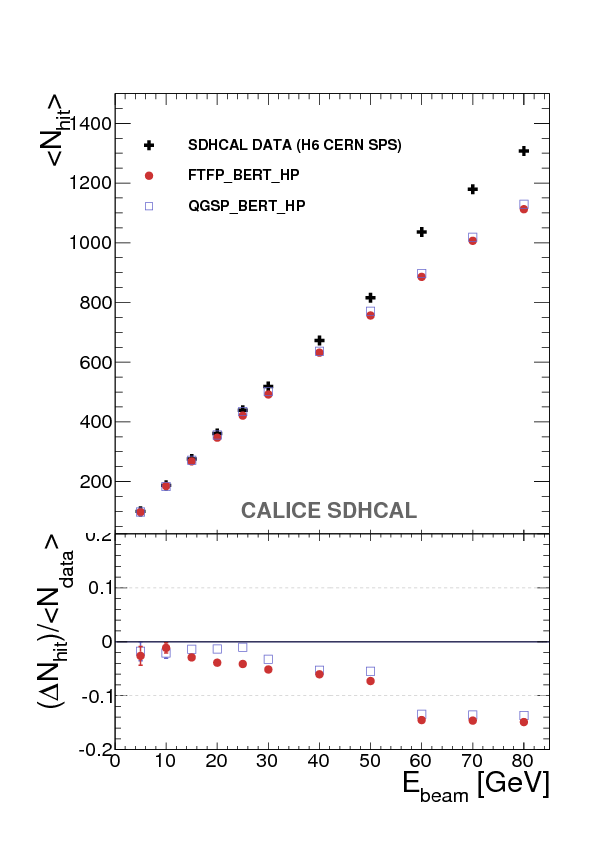}}
\caption{Average number of hits, firing threshold 3 only (left) or any threshold (right), as a function of the beam energy for pion runs. Data are represented by black crosses, simulations are represented by red circles and open blue squares for FTFP\_BERT\_HP and QGSP\_BERT\_HP physics lists respectively. Relative deviations are also presented~\cite{Deng:2016obt}.}
\label{fig:hitvsenergy}
\end{figure}

All digitisation parameters are tuned to the muon data except the effect of charge screening
by 2 avalanches in close vicinity. The avalanche screening modelling is tuned to reproduce 
the mean number of hits in electron showers. With this simulation complete, comparisons of data and simulation can be done as shown on figure~\ref{fig:hitvsenergy}. 
For the mean number of hits above the 3rd threshold inside pion showers, the agreement between data and simulation is good. The 3rd threshold is sensitive to the most dense 
part of the shower, that is the one mostly due to the shower electromagnetic fraction. 
On the contrary, the mean numbers of hits shows a discrepency for pion energies above 30 GeV. Comparisons on shower topological variables done in \cite{Steen:2015wet} indicate that modelling of hadronic showers by GEANT4 might be the source of the observed discrepencies between SDHCAL prototype data and simulation.
    
\section{Summary} 

A prototype of a semi-digital imaging calorimeter has been build
using GRPCs as sensitive detectors. 
The GRPCs are associated with an embedded readout electronics and 
allow a very fine segmentation of the readout of 9216 channels per 
m$^2$. 
The prototype shows good performance in reconstructing the pion energies. 
GRPC electronic response has been modelled in the simulation and the model 
has been tuned to data using muon beams and electron showers. 
This modelling is precise enough to compare data and simulation for pion showers
inside the SDHCAL technological prototype, allowing to test
models of hadronic interactions in matter.

\acknowledgments

This work has been done within the SDHCAL group of the CALICE 
collaboration. The SDHCAL group comprises teams from 
IPNL (Lyon, France), LLR (Palaiseau, France), LAPP (Annecy, France),
LPC (Clermont-Ferrand, France), OMEGA (Palaiseau, France), 
CIEMAT (Madrid, Spain), UCL (Louvain, Belgium),
Universiteit Gent (Belgium), Tsinghua University (Beijing, China), 
Universit\'e de Tunis (Tunisia).


\end{document}